\documentclass[espf]{aa}
\usepackage{graphicx}
\def\kms{km\,s$^{-1}$}
\begin{document}
\title{Multi-site, multi-technique survey of $\gamma$\,Doradus candidates}

   \subtitle{I. Spectroscopic results for 59 stars}

\author{P. Mathias\inst{1} \and
        J.-M. Le Contel\inst{1} \and
        E. Chapellier\inst{1} \and
        S. Jankov\inst{1} \and
        J.-P. Sareyan\inst{1} \and
        E. Poretti\inst{2} \and
        R. Garrido\inst{3} \and
        E. Rodr\'{\i}guez\inst{3} \and
        A. Arellano Ferro\inst{4} \and
        M. Alvarez\inst{5} \and
        L. Parrao\inst{4} \and
        J. Pe\~na\inst{4} \and
        L. Eyer\inst{6} \and
        C. Aerts\inst{7} \and
        P. De Cat\inst{7} \and
        W.W. Weiss\inst{8} \and
        A. Zhou\inst{9}
       }

\offprints{P. Mathias}

\institute{Observatoire de la C\^ote d'Azur, Dpt. Fresnel, UMR 6528,
           F-06304 Nice Cedex 4\\
           \email{mathias@obs-nice.fr} \and
           Osservatorio Astronomico di Brera, Via Bianchi 46,
           I-23807 Merate \and
           Instituto de Astrofisica de Andalucia, Apt. 3044,
           E-18080 Granada \and
           Instituto de Astronomia, Universidad Nacional Autonoma de Mexico,
           Apdo. Postal 70-264, Mexico D.F., 04510 \and
           Observatorio Astron\'omico Nacional, IA-UNAM, Apto.Postal 877,
           Ensenada, B.C., M\'exico, 22800 \and
           Princeton University Observatory, Princeton, NJ 08544, USA \and
           Katholieke Universiteit Leuven, Departement Natuurkunde en Sterrenkunde,
           Celestijnenlaan 200 B, B - 3001 Leuven \and
           Institute for Astronomy, University of Vienna, Tuerkenschanzstrasse 17,
           A-1180 Vienna \and
           National Astronomical Observatories, Chinese Academy of Sciences,
           Beijing 100012, PR China
          }

\date{Received October 15, 2003; accepted ...}

\abstract{
We present the first results of a 2-year high-resolution spectroscopy campaign 
of 59 candidate $\gamma$\,Doradus stars which were mainly discovered 
from the HIPPARCOS astrometric mission.
More than 60\,\% of the stars present line profile variations which can be interpreted
as due to pulsation related to $\gamma$\,Doradus stars.
For all stars we also derived the projected rotation velocity (up to more than
200\,\kms).
The amplitude ratios $2K/\Delta m$ for the main HIPPARCOS frequency are in
the range 35 - 96\,\kms\,mag$^{-1}$.
Less than 50\,\% of the candidates are possible members of binary systems, 
with 20 stars being confirmed $\gamma$\,Doradus.
At least 6 stars present composite spectra, and in all but one case (for
which only one spectrum could be obtained), the narrow component shows line
profile variations, pointing towards an uncomfortable situation if this
narrow component originates from a shell surrounding the star.
This paper is the first of a series concerning mode identification using both
photometric and spectroscopic methods for the confirmed $\gamma$\,Doradus
stars of the present sample. 
         }
\maketitle

\section{Introduction}

In the coming decade, thanks to dedicated satellites (COROT, EDDINGTON), the 
detailed knowledge of the internal structure of stars should be achieved through the 
technique of asteroseismology. 
The goal of this, relatively new, research domain is to derive the internal 
processes in stars with an unprecedented precision through a detailed study of 
their oscillations. 

This paper deals with a class of non-radial pulsators along the main sequence, namely 
the $\gamma$\,Doradus stars (see e.g. Kaye et al. (\cite{kh99a}) for the main
observational characteristics of this class of variables). 
These stars are multiperiodic high-order gravity-mode oscillators with spectral
types around F0.
The origin of the mode destabilization is not clearly known yet, and
driving mechanisms have been proposed by Guzik
et al. (\cite{gk00}), Wu (\cite{w02}) and L\"offler (\cite{l02}). 

Much effort is currently made to find new members of this group, to constrain
their pulsation characteristics and their position in the HR Diagram, especially
the $\gamma$\,Doradus star's red border in relation with the solar-like star's blue
border.
Indeed, they show quite a large variety in their observational behaviour, and
the number of confirmed members is still low.
This observational campaign should contribute to the necessary comparison
between the observational HR diagram and the theoretical one 
recently defined by Warner et al. (\cite{wk03}).
Because of their relatively low amplitude (few tens of mmag in photometry, of the order
of 1\,\kms\ in radial velocity), and due to the long time scales of the variation
(between 0.3 and 3\,d), the detection of such variables is still difficult.
Up to now the best tool has been the HIPPARCOS satellite. 
The HIPPARCOS
sampling does not suffer from the aliasing problems of a single Earth
site which is of particular annoyance for $\gamma$\,Dor studies.
However, there are two major drawbacks: the precision of photometric
individual measurements degrades quite rapidly for fainter stars and
the non-continuous sampling makes the detection/interpretation of
multiperiodic phenomena difficult.

Several studies selected $\gamma$\,Dor candidates from HIPPARCOS: Eyer
(\cite{e98}) proposed a list of such candidates extracted from the periodic 
variable stars
in the HIPPARCOS variability annex which have well defined absolute magnitude
and colour.  Aerts, Eyer and Kestens (1998) used stars from the same
catalogue which have furthermore Geneva photometry.
It permitted to use a multivariate discriminant analysis which
proved to be very efficient for detecting new slowly pulsating B stars (Waelkens
et al.\ \cite{wa98}), which are also main-sequence gravity-mode oscillators.
Handler (\cite{h99}) broadened the search for $\gamma$\,Dor stars to the
unsolved variable stars of the HIPPARCOS variability annex and relaxed
selection criteria, focusing more on the nature of the power spectra. 
These studies proposed about 60 bona fide and prime candidates stars.
One star in our sample, \object{HD\,173977}, which was in Handler's list 
(\cite{h99}), has been discarded since it is now classified as a $\delta$\,Scuti
variable (Chapellier et al. \cite{cm03}).

However, the spectroscopic studies of most of the candidates having
well-known photometric properties are much less detailed. 
In 2001, we undertook
a spectroscopic campaign whose
objective was twofold:
\begin{itemize}
\item to derive basic spectroscopic parameters (rotation velocity, line
profile variations, duplicity, etc.) for better identification of the
pulsation modes. This is particularly important for stars with similar
rotation and pulsation frequencies
(Dintrans \& Rieutord \cite{dr00}).
\item to prepare the COROT and EDDINGTON space missions by including
at least one $\gamma$\,Doradus star in the core program (Mathias et al.
\cite{mc03a}, \cite{mc03b}). 
\end{itemize}

This paper presents the first descriptive part of the campaign concerning
59 $\gamma$\,Doradus candidates.
Only the very homogeneous OHP spectroscopy is discussed.
The observations are described in
Sect.\,2.
In Sect.\,3, results are given for individual stars, depending on the detection
of line profile variations, in Sect.\,3.
Sect.\,4 and 5 present the pulsation and stellar
environments for some candidates.
Concluding remarks are given in Sect.\,6.

\section{The data}

The spectroscopic data were obtained at the Observatoire de Haute-Provence,
using the AURELIE spectrograph attached to the 1.52\,m telescope.
The spectral domain covers the range 4470--4540\AA\ with a resolution
power $\lambda/\Delta\lambda=55\,000$, which is enough to detect the
expected low degree Line Profile Variations (LPV).
We focused our study on the two unblended lines of \ion{Fe}{ii} and \ion{Ti}{ii}
at respectively 4501.273 and 4508.288\,\AA\ .
Spectra were reduced using the standard packages of IRAF.
The exposure time was adapted to ensure a S/N ratio above 150, but limited to
1\,h to avoid phase smearing.

The selected targets and the main characteristics of the observations 
are presented in Table\,\ref{table1} and Table\,\ref{table1bis}.
Table\,\ref{table1} summarizes the observational results on the stars
showing LPV, which represents 2/3 of the sample.  We report below 
on some special cases. 
Table\,\ref{table1bis} lists the stars for which nothing has been detected, 
which does not mean that LPV are not present (LPV below our detection
threshold, or insufficient data); in some cases spectra are just
useful to detect accurate $v\,\sin i$ values.  
\begin{table*}
\caption[]{$\gamma$\,Doradus stars candidates observed at OHP
showing LPV.  The first 2 columns
give the HD or HIP star number (if this latter exists).
Asterisked stars are both good $\gamma$\,Doradus candidates and
possible members of binary systems (see Sect.\,5).
Next columns respectively
provide the number of spectra, the observation window, the mean
exposure time, the mean S/N value, the number of photometric
frequencies previously detected, the LPV signature, 
remarks about the binarity, the $v\,\sin i$ value and some references$^{*}$.}
\begin{center}
\begin{tabular}{rrrr rr cl ll cc}
\hline
\multicolumn{1}{c}{HD} &
\multicolumn{1}{c}{HIP} &
\multicolumn{1}{c}{$N$} &
\multicolumn{1}{c}{Range} &
\multicolumn{1}{c}{Exp.} &
\multicolumn{1}{c}{S/N} &
\multicolumn{1}{c}{Phot.} &
\multicolumn{1}{c}{LPV} &
\multicolumn{1}{c}{Binarity} &
\multicolumn{1}{c}{$v\,\sin i$} &
\multicolumn{1}{c}{Ref} \\
\multicolumn{1}{c}{} &
\multicolumn{1}{c}{} &
\multicolumn{1}{c}{} &
\multicolumn{1}{c}{[d]} &
\multicolumn{1}{c}{[sec]} &
\multicolumn{1}{c}{} &
\multicolumn{1}{c}{} &
\multicolumn{1}{c}{} &
\multicolumn{1}{c}{} &
\multicolumn{1}{c}{[\kms]} &
\multicolumn{1}{c}{} \\
\hline
\multicolumn{11}{c}{\it Bona fide $\gamma$ Dor stars}\\
\noalign{\smallskip}
   277 &   623      & 24 & 527 & 3500 & 215 &  3 & strong&  &  31     & H99 HF01 FW03  \\
 12901 &  9807      &  5 & 268 & 2000 & 240 & 3 & evident         &             & 66      & H99 EA00 AC03 \\
 62454 & 37863      & 12 & 353 & 2800 & 240 & 5 & in the primary & SB2 & 10.5 and 5 &     KH99   \\
 68192$^{*}$ & 40462      &  3 &   4 & 2800 & 220 & 2 & weak &  RV var. & 95 &   HE99 KH99  \\
 86371$^{*}$ & 48830      &  4 & 348 & 1900 & 220 & 2 & weak ? & SB2, ell. ? &  11 and 6  & H99 KE02 \\
105458$^{*}$ & 59203      & 25 & 421 & 3600 & 200 & 6 & evident & binary? &   39    &  H99 HF01   \\
108100$^{*}$ & 60571      &  5 & 322 & 2700 & 190 & 3 & in the slow rot.& binary &  65 and 13 & BH97 HF02 \\
164615 & 88272      & 13 & 379 & 2300 & 180 & 4 &evident &  &   65    & ZR97 E98  \\
167858$^{*}$ & 89601      & 11 & 366 & 2200 & 210 & 2 & evident but small & SB &   9.0  & E98 AE98 H99 FW03\\
218396 &114189      &162 & 505 & 1100 & 250 & 4 & evident & &   38    & E98 ZR99  \\
221866$^{*}$ &116434  & 32 & 506 & 3300 & 220 &   2 & in the sec. & SB2 & 13 and 10  &H99 FW03\\
224638 &118293      & 23 & 410 & 3300 & 200 & 5 & evident & &   17    & PK02\\
224945 &   159      &  2 & 346 & 1800 & 190 & 5 &         & &   54    & PK02 \\
\multicolumn{11}{c}{\it Prime candidates}\\
\noalign{\smallskip}
  2842$^{*}$ &  2510      &  2 & 346 & 3600 & 210 & 3 & in the blue wing      & maybe     &   77  & H99 KE02 FW03  \\
  7169$^{*}$ &  5674      &  4 & 137 & 3200 & 210 & 2 & in both stars& visual &90 and 8 &  H99 FW03  \\
  9365$^{*}$ &  7280      &  6 & 269 & 4000 & 200 & 2 & weak            & RV var. & 69     & H99 FW03 LJ89 \\
 23874$^{*}$ & 17826      &  4 &   5 & 3600 & 200 & 2 & in the slow rot.& visual &90 and 9 & H99 FW03 \\
 48271 & 32263      & 19 & 423 & 3400 & 240 & 6 & evident         &             & 21   & E98 H99 MB03\\
 63436$^{*}$ & 38138      &  7 & 347 & 3000 & 190 & 2 & in the blue wing & binary ?   &   66   & H99 MB03  \\
 70645$^{*}$ & 41488      & 10 & 423 & 3600 & 230 & 2 & evident & SB1 &   11   & H99 MB03  \\
 80731$^{*}$ & 46099      & 11 & 422 & 3400 & 210 & 5 & evident &SB1 &   13   &  H99 MB03   \\
 86358$^{*}$ & 48895      &  3 &   2 & 1800 & 220 & 2 & in the slow rot. & binary &  37 and 25 &   H99 KE02 FW03  \\
100215$^{*}$ & 56275      & 11 & 422 & 3000 & 240 & 2 & evident & SB1 &   13    &   H99 KE02 FW03  \\
113867$^{*}$ & 63951      &  7 & 572 & 1800 & 210 & 2 & weak & SB2 ? &  8.5 and 110&  H99 FW03  \\
171244$^{*}$ & 90919      &  4 & 316 & 3000 & 200 & 2 & very weak in the wings &binary? & 50 &H99 FW03 \\
175337 & 92837      &  7 & 315 & 3300 & 230 & 2  &  evident & unlikely &  38    &  H99 KE02 FW03 \\
195068 &100859      & 36 & 564 & 1300 & 240 & 2 &evident &  &  46    &  E98 H99 FW03  \\
211699 &110163      & 21 & 399 & 3700 & 210 & 2 & evident & & 12    &  H99  \\
\multicolumn{11}{c}{\it Other candidates and COROT targets}\\
\noalign{\smallskip}
 44195 & 30154      &  4 & 345 & 3000 & 220 & cst & weak ?         &  &  50   & PG02  \\
 44333 & 30217      &  2 &   3 & 1400 & 160 &     & in the slow rot.?& binary &160 and 15& PG02 RG02  \\
 49434 & 32617      & 92 & 422 &  850 & 230 &  & in the blue wing   &          &   82   & BC02  \\
112429 & 63076      & 40 & 566 &  900 & 220 &2& in the wings &   & 101    &  E98 AE98 FW03 \\
171834 & 91237      & 13 &  11 &  940 & 200 & cst & very weak & &  65 &  GP02 \\
171836 & 91272      &  3 &  75 & 3600 & 200 & cst & observed &  &  62    &  GP02 \\
172506 & 91580      &  2 &   9 & 3100 & 180 & var? & observed &binary ? &  39    &  GP02  \\
174353 & -          &  2 &   8 & 3600 & 200 &    & varying res. flux & & 11    & H02 \\
\hline
\end{tabular}
\end{center}
\label{table1}
\begin{list}{}{}
\item[$^{*}$] AC03: Aerts et al. \cite{ac03} - 
AE98: Aerts et al. \cite{ae98} - 
BC02: Bruntt et al. \cite{bc02} - 
BH97: Breger et al. \cite{bh97} - 
E98: Eyer \cite{e98} - 
EA00: Eyer \& Aerts \cite{ea00} - 
FW03: Fekel et al. \cite{fw03} - 
GP02: Garrido et al. \cite{gp02} - 
H99: Handler \cite{h99} - 
H02: Handler \cite{h02} - 
HE99: Henry \cite{he99} -
HF01: Henry et al \cite{hf01} -
HF02: Henry \& Fekel \cite{hf02} - 
KE02: Koen \& Eyer \cite{ke02} -
KH99: Kaye et al. \cite{kh99b} -
LJ89: Liu et al. \cite{lj89} -
MB03: Mart\'{\i}n et al. \cite{mb03} -
PG02: Poretti et al. \cite{pg02} -
PK02: Poretti et al. \cite{pk02} -
RG02: Royer et al. \cite{rg02} -
ZR97: Zerbi et al. \cite{zr97} -
ZR99: Zerbi et al. \cite{zr99}.
\end{list}
\end{table*}
\begin{table*}
\caption[]{$\gamma$\,Doradus stars candidates observed at OHP
not showing LPV.  The first 2 columns
give the HD or HIP star number (if this latter exists).
Next columns respectively
provide the number of spectra, the observation window, the mean
exposure time, the mean S/N value, the status
of their $\gamma$ Dor variability (BF: Bona fide; PC: prime
candidate; COROT: possible COROT target chosen as a $\gamma$\,Doradus candidate
from its spectral type), the number of photometric
frequencies previously detected, general remarks,
the $v\,\sin i$ value and some references$^{*}$.}
\begin{center}
\begin{tabular}{rrrr rr cc  lr c}
\hline
\multicolumn{1}{c}{HD} &
\multicolumn{1}{c}{HIP} &
\multicolumn{1}{c}{$N$} &
\multicolumn{1}{c}{Range} &
\multicolumn{1}{c}{Exp.} &
\multicolumn{1}{c}{S/N} &
\multicolumn{1}{c}{status} &
\multicolumn{1}{c}{N} &
\multicolumn{1}{c}{Remarks} &
\multicolumn{1}{c}{$v\,\sin i$} &
\multicolumn{1}{c}{Ref} \\
\multicolumn{1}{c}{} &
\multicolumn{1}{c}{} &
\multicolumn{1}{c}{} &
\multicolumn{1}{c}{[d]} &
\multicolumn{1}{c}{[sec]} &
\multicolumn{1}{c}{} &
\multicolumn{1}{c}{} &
\multicolumn{1}{c}{} &
\multicolumn{1}{c}{} &
\multicolumn{1}{c}{[\kms]} &
\multicolumn{1}{c}{} \\
\hline
 40745 & 28434      &  2 &   5 & 1500 & 210 & PC   &   2   &          & 37   & E98 AE98 \\
 41448 & 28778      &  2 &   5 & 2800 & 160 & PC   &   2   &          & 93   & E98 H99 \\
 43338 & 29758      &  5 & 346 & 2900 & 230 & COROT& cst   &          & 170   &PG03 \\
 44716 & -          &  2 &   9 & 3600 & 220 & COROT& geom.?& binary?  & 10   & PG03  \\
 45138 & -          &  2 &  52 & 3600 & 140 & COROT&  cst  &          & 60   & PG03  \\
 45196 & 30611      &  2 &  48 & 3600 & 200 & COROT& var   & $\delta$ Sct?&190   & PG03  \\
 46304 & 31167      & 23 &   7 & 1100 & 150 & COROT& undetectable & binary ?  &  190   & PG02 RG02 \\
 56359 & 35248      &  1 &   0 & 2100 & 170 & COROT& var   &          &210   & PG03  \\
 65526 & 39017      &  1 &   0 & 3600 & 190 & BF   & 4     &          & 59   & H99 MB03  \\
 69715$^{*}$ & 40791      &  2 &   4 & 2700 & 210 & PC   & 4     & binary?  & 145   & H99 MB03 \\
152569 & 82693      &  2 & 368 & 1800 & 190 & BF   &       & $\delta$ Sct& 175    & KH00  \\
152896$^{*}$ & 82779      &  2 & 318 & 2400 & 230 & PC   & 2     & SB2      &     & H99 FW03  \\
155154$^{*}$ & 83317      &  3 & 348 & 1200 & 220 & BF   & 4     & binary?  & 175    & H99 HF01 \\
160295 & 86374      &  1 &   0 & 1800 & 180 & PC   & 2     & binary   &60 and 9   & H99 FW03  \\
164259 & 88175      & 23 &  43 &  600 & 170 & COROT&       &        &  72 & LL01  \\
165645 & 88565      &  3 & 317 & 2700 & 170 & BF   &       &        &  128& K98  \\
172423 & -          &  2 &   9 & 3600 & 190 & COROT&       &        & 14  & H02 \\
173073 & 91838      &  3 &  75 & 3600 & 190 & COROT&cst    &        & 62  & GP02 \\
174704 & -          &  2 &  10 & 3600 & 220 & COROT&       &        & 9.0 & H02 \\
175431 & -          &  2 &   8 & 3700 & 190 & COROT&       &        & 62  & H02 \\
178596 & 94068      &  2 &  77 & 1500 & 200 & COROT&       &        & 68  & H02 \\
184064$^{*}$ & -    &  2 &   4 & 3600 & 230 & COROT&       & SB ?   & 9.3 & H02 \\
206043 &106897      &  3 & 312 &  620 & 200 & BF   & 3     &        &120  & E98 HF01 FW03\\
\hline
\end{tabular}
\end{center}
\label{table1bis}
\begin{list}{}{}
\item[$^{*}$] 
AE98: Aerts et al. \cite{ae98} - 
E98: Eyer \cite{e98} - 
FW03: Fekel et al. \cite{fw03} - 
GP02: Garrido et al. \cite{gp02} - 
H99: Handler \cite{h99} - 
H02: Handler \cite{h02} - 
HF01: Henry et al \cite{hf01} -
HF02: Henry \& Fekel \cite{hf02} - 
K98: Kaye \cite{k98} - 
KH00: Kaye et al. \cite{kh00} -
LL01: Lastennet et al. \cite{ll01} -
MB03: Mart\'{\i}n et al. \cite{mb03} -
PG02: Poretti et al. \cite{pg02} -
PG03: Poretti et al. \cite{pa03} -
RG02: Royer et al. \cite{rg02} -
\end{list}
\end{table*}
The projected rotation velocities $v\,\sin i$ are computed as the mean value of the first
zero of the Fourier transform corresponding to the two considered unblended
lines.
The uncertainties are typically of 5\,\kms\ for rapid rotators
and of the order of 1\,\kms\ for the slow ones.
The radial velocities were obtained with two methods: a simple Gaussian fit
and first line moment.
Both methods have their limitations: for the first one, it is accurate only
when LPV are small, while the second one presents larger errors due to
the uncertainty on the location of the integration domain, and a stronger
effect of the noise.
Generally, only the first moment was computed when the line profiles were too 
broad i.e., when the projected rotation velocity value was above 50\,\kms.

\section{Results on stars with marked or suspected LPV}

\begin{figure}
\centering
\includegraphics[width=8cm]{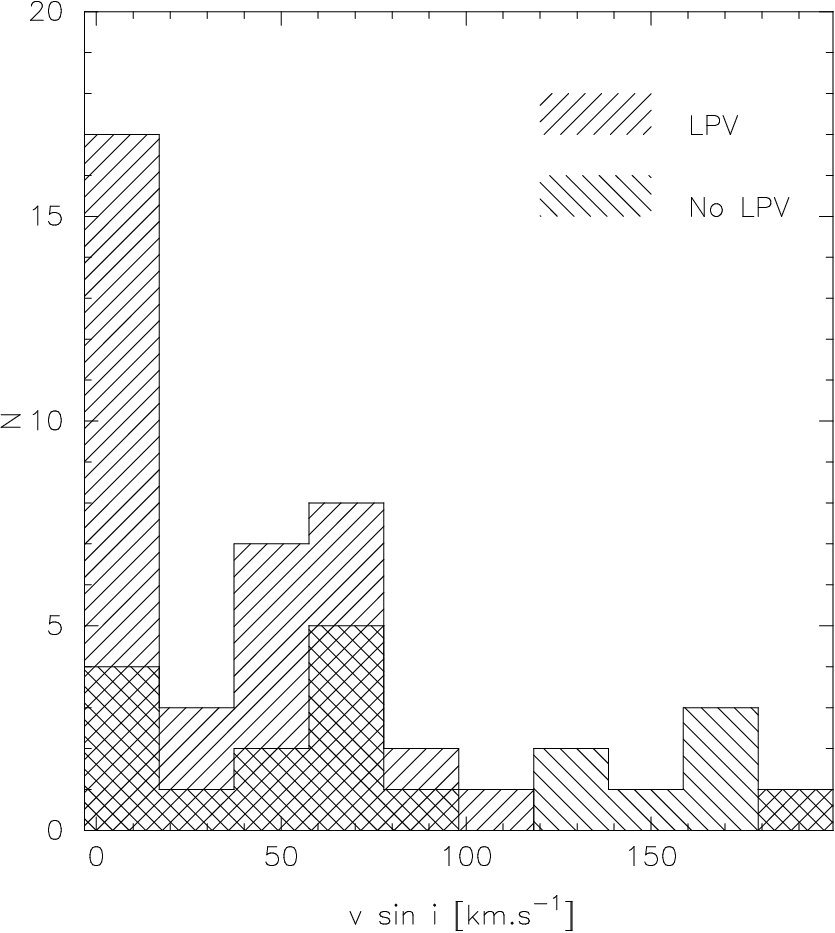}
\caption{Histogram showing the number of stars $N$ showing LPV 
as a function of the projected rotation velocity.
It can be noticed that LPV are easier to detect for stars having low
$v\,\sin i$ values.
}
\label{rotlpv}
\end{figure}

Most of the 59 stars of our sample were chosen in the updated web  
list\footnote{http://www.astro.ac.at/~dsn/gerald/gdorlist.html} 
initiated by Handler \& Krisciunas (\cite{hk97}),
or in few other works, such as the potential candidates
in the COROT fields.
Therefore, most candidates present light variations, and the objective
was thus to see if LPV were present.
Depending on a few factors, the major ones being the too low S/N
level, the large projected rotation velocity, or too few spectra, sometimes we
could hardly detect LPV.
Even if LPV are easier to detect when $v\,\sin i$
is small, they were also detected in some rapid rotators. 
As shown in Fig.\,\ref{rotlpv}, setting a limit of 80\,\kms, we find 
35 stars with LPV against 12 stars without LPV below the detection threshold,
and 4 stars with LPV against 8 stars without LPV above this limit.
Even for those targets that do show LPV, it is not always clear
if the variations are caused by $g$-modes or not.
This is due to unexpected line profile behaviour 
(such as a change of the residual flux in the line core) or
to the lack of clear periodicity in the observed LPV.
However, we can be confident that when more than one period is found
in photometry, and when LPV are present, the star can be considered
as a real $\gamma$\,Doradus star.
This is the case for about 40\,\% of the stars in our sample.

\begin{figure}
\centering
\includegraphics[width=8cm]{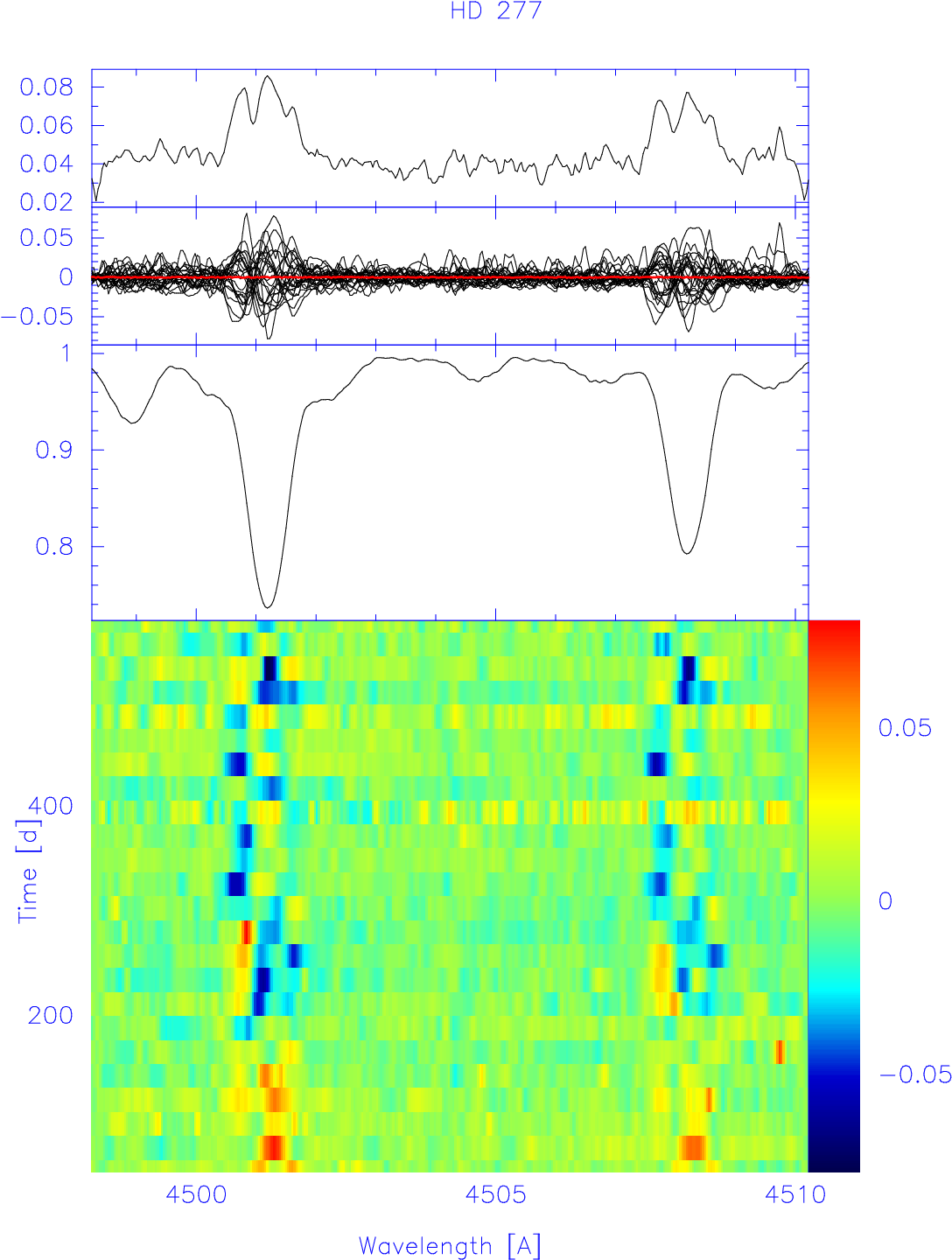}
\caption{Lower part: Bi-dimensional plot of the temporal evolution 
(on an arbitrary scale) of the 
residual spectra for the star HD\,277. 
Upper part, from bottom to top are successively represented the mean
spectrum, the individual residual spectra and the dispersion $\sigma$ around 
the mean residual (see text).
}
\label{figure1}
\end{figure}

The detection of LPV has been done using visual inspection of the
profiles of the two considered unblended lines. 
The way the profile is distorted was generally the same for both ions.
Also, we computed the standard deviation of individual spectra 
with respect to the average one. 
If present, LPV should manifest as an increase of this
deviation along the line.
Moreover, for $g$-modes, the ratio of tangential over radial velocity
is large, so line wings should be more perturbed than they are for
$p$-modes.
This particular form is called hereafter a typical $g$-mode behaviour.
Figure \,\ref{figure1} provides an example of the application of
these tools to the spectra of \object{HD\,277}.
We also note
that when LPV are not particularly prominent, the presence of an undetected 
companion can be responsible for the modifications of the line wings 
(\object{HD\,2842}, \object{HD\,49434}, \object{HD\,63436}, 
\object{HD\,112429} and \object{HD\,171244}.
For the first 3 stars in particular, only the blue wing is perturbed).

In the following, we discuss in detail some special cases, i.e., 
spectroscopic binaries
with two spectra (\object{HD\,62454},\object{HD\,86371}, \object{HD\,221866}),
with one spectrum only (\object{HD\,70645}, \object{HD\,80731}, 
\object{HD\,100215}), and unclear cases of spectroscopic peculiarities 
(\object{HD\,108100}, \object{HD\,113867}, \object{HD\,211699}).

\subsection{\object{HD\,62454}} 

\begin{figure}
\centering
\includegraphics[width=8cm]{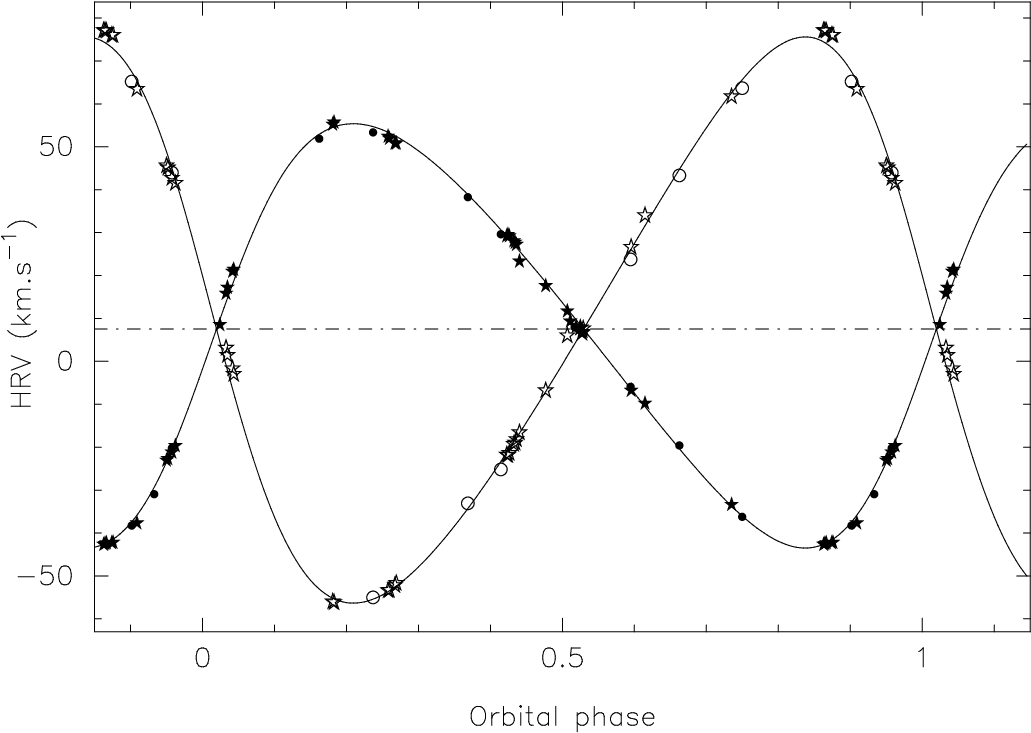}
\caption{Fit of the orbit of HD\,62454.  
Empty symbols represent the companion velocities.
The values of Kaye et al. (1999) are represented as stars.
The dot-dash line represents the heliocentric
velocity of the system.
}
\label{figure2}
\end{figure}

This is a $\gamma$\,Doradus star first proposed by Henry 
(\cite{he99}).
Kaye et al. (\cite{kh99a}) discovered that it is a double-lined spectroscopic
binary with a 11.6\,d orbital period.
To increase the accuracy of the orbital parameters, we computed a new
ephemeris by including their velocities in addition to ours.
The deduced orbital elements are given in Table\,\ref{table2}, and the binary
motion in Fig\,\ref{figure2}.
\begin{table}
\caption[]{Parameters of the binary orbit for the SB2 star HD\,62454.
Our data together with those of Kaye et al. (1999) lead to a final rms of the 
residuals of 1.81\,\kms.
}
\begin{center}
\begin{tabular}{rcl}
\hline
$P$            & = & $11.61550 \pm 0.00015$\,d    \\
$T_0$          & = & $2450996.55 \pm 0.04$\,d  \\
$e$            & = & $0.205 \pm 0.005$           \\
$\gamma$       & = & $7.5 \pm 0.2$\,\kms \\
$K_1$          & = & $49.4 \pm 0.4$\,\kms \\
$\omega_1$     & = & $261.0 \pm 1.3^{\circ}$     \\
$K_2$          & = & $66.0 \pm 0.4$\,\kms \\
$\omega_2$     & = & $81.0 \pm 1.3^{\circ}$     \\
$a_1 \sin i$   & = & $(7.73 \pm 0.07)\,10^6$\,km   \\
$M_1.\sin^3 i$ & = & $0.994 \pm 0.013$\,$M_{\sun}$ \\
$a_2 \sin i$   & = & $(10.31 \pm 0.08)\,10^6$\,km   \\
$M_2.\sin^3 i$ & = & $0.745 \pm 0.011$\,$M_{\sun}$ \\
\hline
\end{tabular}
\end{center}
\label{table2}
\end{table}
Kaye et al. (\cite{kh99b}) were able to detect up to 5 frequencies in their photometric 
data, implying that this star is a confirmed $\gamma$\,Doradus star.
After removing the orbital motion and spectra containing the component lines, the
primary indeed presents LPV.
More observations will be necessary to observe the pulsation radial
velocities.
The projected rotation velocity is 10.5\,\kms, a value comparable to the
11.5\,\kms\ determined by Kaye et al. (\cite{kh99b}), while we confirm the
5\,\kms\ value of the $v\,\sin i$ of the companion.

\subsection{\object{HD\,70645}} 

\begin{figure}
\centering
\includegraphics[width=8cm]{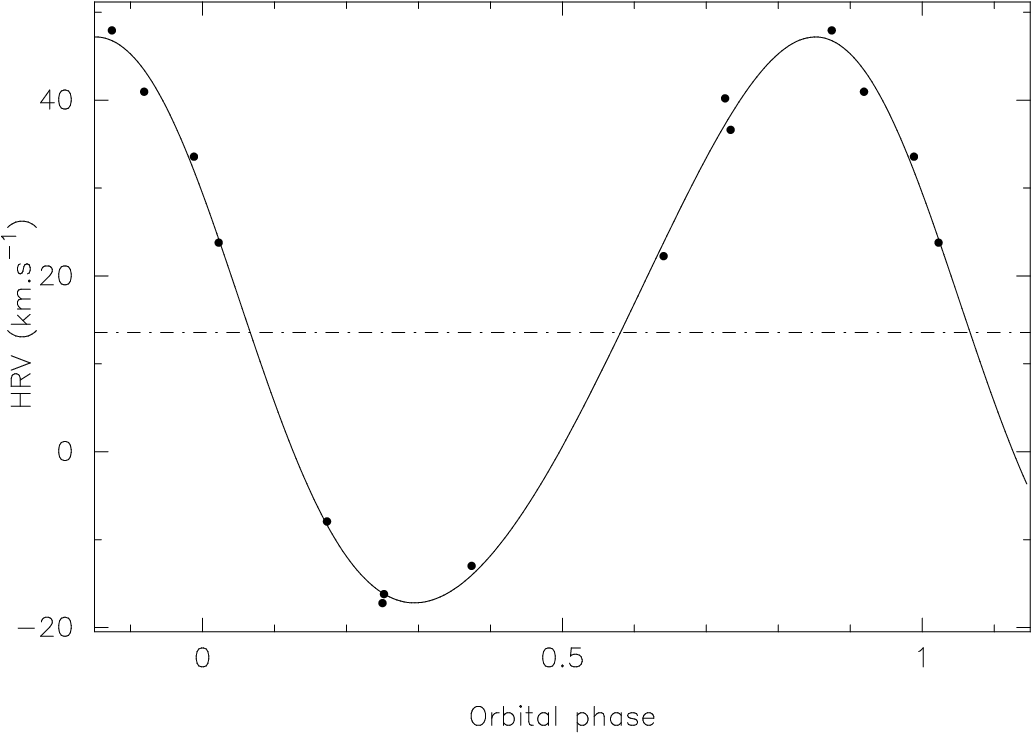}
\caption{Fit of the orbit of HD\,70645. The dot-dash line represents the heliocentric
velocity of the system.
}
\label{figure3}
\end{figure}

Handler (\cite{h99}) classified this star as a prime 
$\gamma$\,Doradus candidate, with 2 frequencies.
Mart\'{\i}n et al. (\cite{mb03}) also detected 2 frequencies in both HIPPARCOS and
Str\"omgren photometry data.
We discovered that this star is a single-lined spectroscopic binary, whose orbital
elements are provided in Table\,\ref{table3} and the orbit in
Fig\,\ref{figure3}.
\begin{table}
\caption[]{Parameters of the binary orbit for the star HD\,70645.
The final rms of the residuals is 2.32\,\kms.
Note that we have only used here the velocities computed from a Gaussian fit.}
\begin{center}
\begin{tabular}{rcl}
\hline
$P$          & = & $8.4450 \pm 0.0025$\,d    \\
$T_0$        & = & $2452270.82 \pm 0.52$\,d  \\
$e$          & = & $0.100 \pm 0.033$           \\
$\gamma$     & = & $13.6 \pm 0.8$\,\kms \\
$K$          & = & $32.2 \pm 1.0$\,\kms \\
$\omega$     & = & $63. \pm 22.^{\circ}$     \\
$a \sin i$   & = & $(3.72 \pm 0.13)\,10^6$\,km   \\
$M.\sin^3 i$ & = & $0.029 \pm 0.003$\,$M_{\sun}$ \\
\hline
\end{tabular}
\end{center}
\label{table3}
\end{table}
After removing the orbital motion, we derived a projected rotation velocity of 
11\,\kms.
More spectroscopic observations are necessary to derive the pulsation 
radial velocities.
LPV are easily detected, and this star should be definitively classified as a 
bona fide $\gamma$\,Doradus star.

\subsection{\object{HD\,80731}} 

\begin{figure}
\centering
\includegraphics[width=8cm]{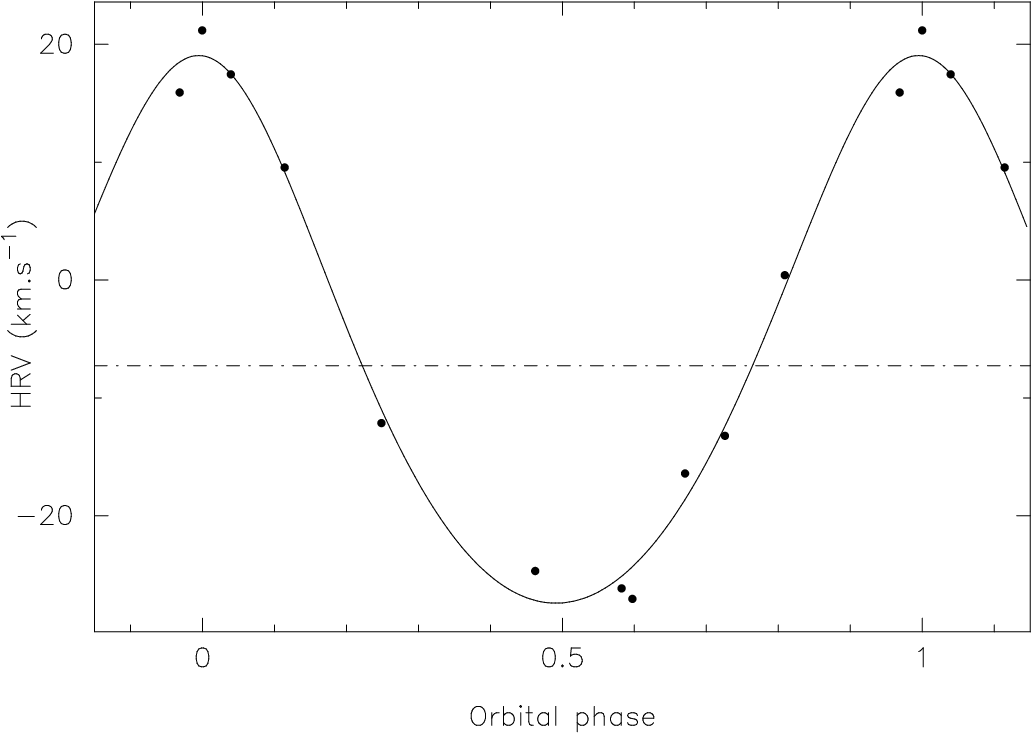}
\caption{Fit of the orbit of HD\,80731. The dot-dash line represents the heliocentric
velocity of the system.
}
\label{figure4}
\end{figure}

This is also a prime $\gamma$\,Doradus candidate detected by
Handler (\cite{h99}) with 2 frequencies.
Mart\'{\i}n et al. (\cite{mb03}) were able to detect up to 5 frequencies in HIPPARCOS and
Str\"omgren photometry data.
We found that this star actually belongs to a binary system, 
and the orbital parameters we derived are
given in Table\,\ref{table4}. The corresponding motion is represented in
Fig.\,\ref{figure4}.
\begin{table}
\caption[]{Parameters of the binary orbit for the star HD\,80731.
The final rms of the residuals is 2.61\,\kms.
Note that we have only used here the velocities computed from a Gaussian fit.}
\begin{center}
\begin{tabular}{rcl}
\hline
$P$          & = & $13.572 \pm 0.011$\,d    \\
$T_0$        & = & $2452283.30 \pm 1.43$\,d  \\
$e$          & = & $0.133 \pm 0.047$           \\
$\gamma$     & = & $-7.3 \pm 1.1$\,\kms \\
$K$          & = & $23.2 \pm 1.2$\,\kms \\
$\omega$     & = & $2. \pm 35.^{\circ}$     \\
$a \sin i$   & = & $(4.30 \pm 0.25)\,10^6$\,km   \\
$M.\sin^3 i$ & = & $0.017 \pm 0.003$\,$M_{\sun}$ \\
\hline
\end{tabular}
\end{center}
\label{table4}
\end{table}
Once the orbital motion is removed, we are able to derive $v\,\sin i=13$\,\kms.
LPV are also present, which manifest mainly as a change in the intensity
level of about 10\,\%, a value well above the 1\,\% uncertainty of the 
continuum level.
This is also visible on the variations of the EW which changes in the same
time by a factor of 2.
This behaviour is
close to that pointed out in \object{HD\,211699} (see below).
We note that this change is not correlated to the binary motion.
More spectroscopic observations are necessary to clearly derive the velocities
associated with the pulsation.

\subsection{\object{HD\,86371}}

This star is considered as a prime candidate by Handler
(\cite{h99}) who derived 2 frequencies from the HIPPARCOS data.
However, Koen \& Eyer (\cite{ke02}) derived, still from HIPPARCOS, a frequency
typical of $\delta$\,Scuti variations (11.68919\,d$^{-1}$).
Using Johnson's photometry, Handler \& Shobbrook (\cite{hs02}) confirm the
$\gamma$\,Doradus character from the amplitude ratio of the $B$ and $V$ filter.
Our 4 spectra show that this star is actually a double-lined spectroscopic binary
with two very similar components, hence very close spectral types, having  projected 
rotation velocities of 6 and 11\,\kms\ respectively.
Therefore, the value of 18\,\kms\ proposed by Royer et al. (\cite{rg02}) is
certainly due to an unfortunate  observing epoch, when the two spectra were 
heavily blended, with radial velocities close to the $\gamma$-value of
the system.
The maximum separation in our data is about 30\,\kms, the minimum
value of the 2K-amplitude of the binary motion.
It is impossible to derive an orbital period, but based on the evolution of
2 consecutive spectra, we estimate a value of around 6 days i.e., close
to twice the shorter frequency derived by Handler (\cite{h99}).
Therefore, this star can be ellipsoidal, but more observations are needed
to derive the orbital ephemeris.
LPV seem to be present as a change in the residual flux of one profile with
respect to the other.
The interpretation of such a change needs to be confirmed from new
observations.

\subsection{\object{HD\,100215}} 

\begin{figure}
\centering
\includegraphics[width=8cm]{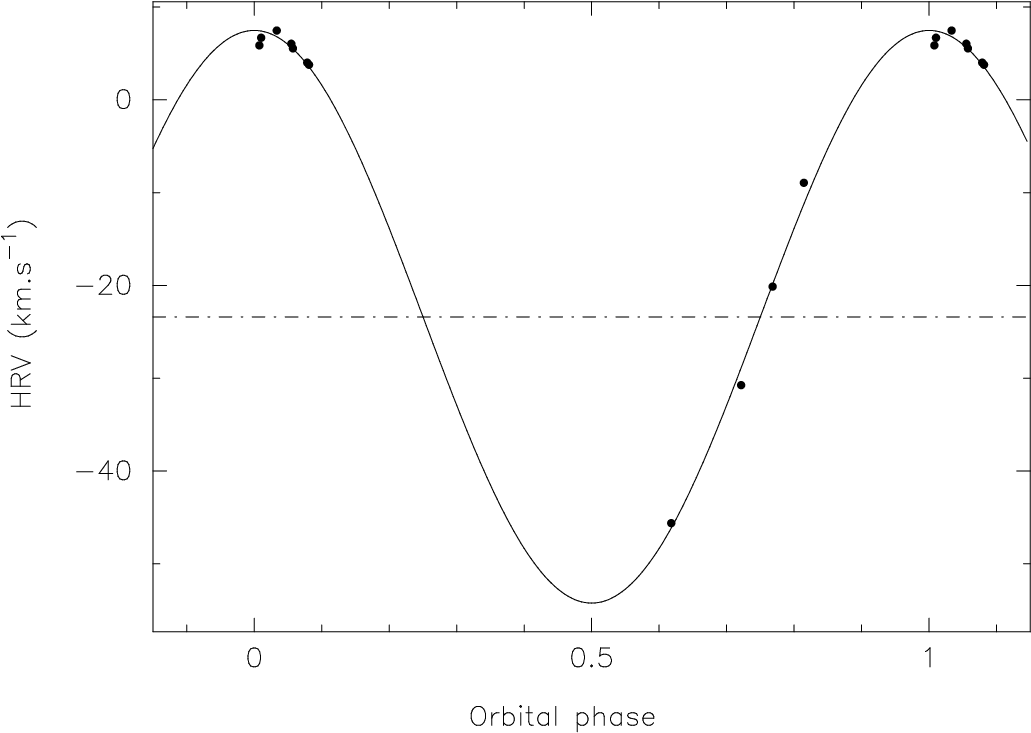}
\caption{Fit of the orbit of HD\,100215 with a fixed null eccentricity 
(see text). 
The dot-dash line represents the heliocentric
velocity of the system.
}
\label{figure5}
\end{figure}

Handler (\cite{h99}) classified this star as a 
prime $\gamma$\,Doradus candidate, with 2 frequencies, the first 
confirmed by Koen \& Eyer (\cite{ke02}).
Fekel et al. (\cite{fw03}) indicate that the star is a member of a binary system
on the basis of the different values of the radial velocities provided in the
literature.
On one of their two spectra, they were able to partially resolve two lines.
Although our data confirm the binarity, 
we were unable to detect a second line system in any of our 11 spectra, 
spread over 422\,d.
Unfortunately, our data are not well sampled to derive a definitive ephemeris.
To converge, we arbitrarily fixed the eccentricity to a null value.
The deduced orbital elements are given in Table\,\ref{table5} and the orbital
motion is represented in Fig.\,\ref{figure5}.
\begin{table}
\caption[]{Parameters of the binary orbit for the star HD\,100215.
Because we have too scarce data, the eccentricity has been arbitrarily
set to zero.
Consequently, the longitude of the periastron is not given.
The final rms of the residuals is 1.35\,\kms.
Note that we have only used here the velocities computed from a Gaussian fit.}
\begin{center}
\begin{tabular}{rcl}
\hline
$P$          & = & $42.628 \pm 0.053$\,d    \\
$T_0$        & = & $2452419.41 \pm 0.29$\,d  \\
$\gamma$     & = & $-23.4 \pm 1.0$\,\kms \\
$K$          & = & $30.9 \pm 1.5$\,\kms \\
$a \sin i$   & = & $(18.09 \pm 0.92)\,10^6$\,km   \\
$M.\sin^3 i$ & = & $0.130 \pm 0.019$\,$M_{\sun}$ \\
\hline
\end{tabular}
\end{center}
\label{table5}
\end{table}
Our value of the projected rotation velocity, 13\,\kms, has to be
compared with the 25\,\kms derived by Fekel et al. (\cite{fw03}).
Since this star presents strong LPV, with an obvious  $\gamma$\,Doradus character, it
is probable that there has been a confusion in Fekel et al.'s interpretation, 
as the travelling bumps at a given phase
produce a profile similar to a double line.
Therefore, this star is clearly confirmed as a new $\gamma$\,Doradus star in a SB1
system.

\subsection{\object{HD\,108100}} 

\begin{figure}
\centering
\includegraphics[width=8cm]{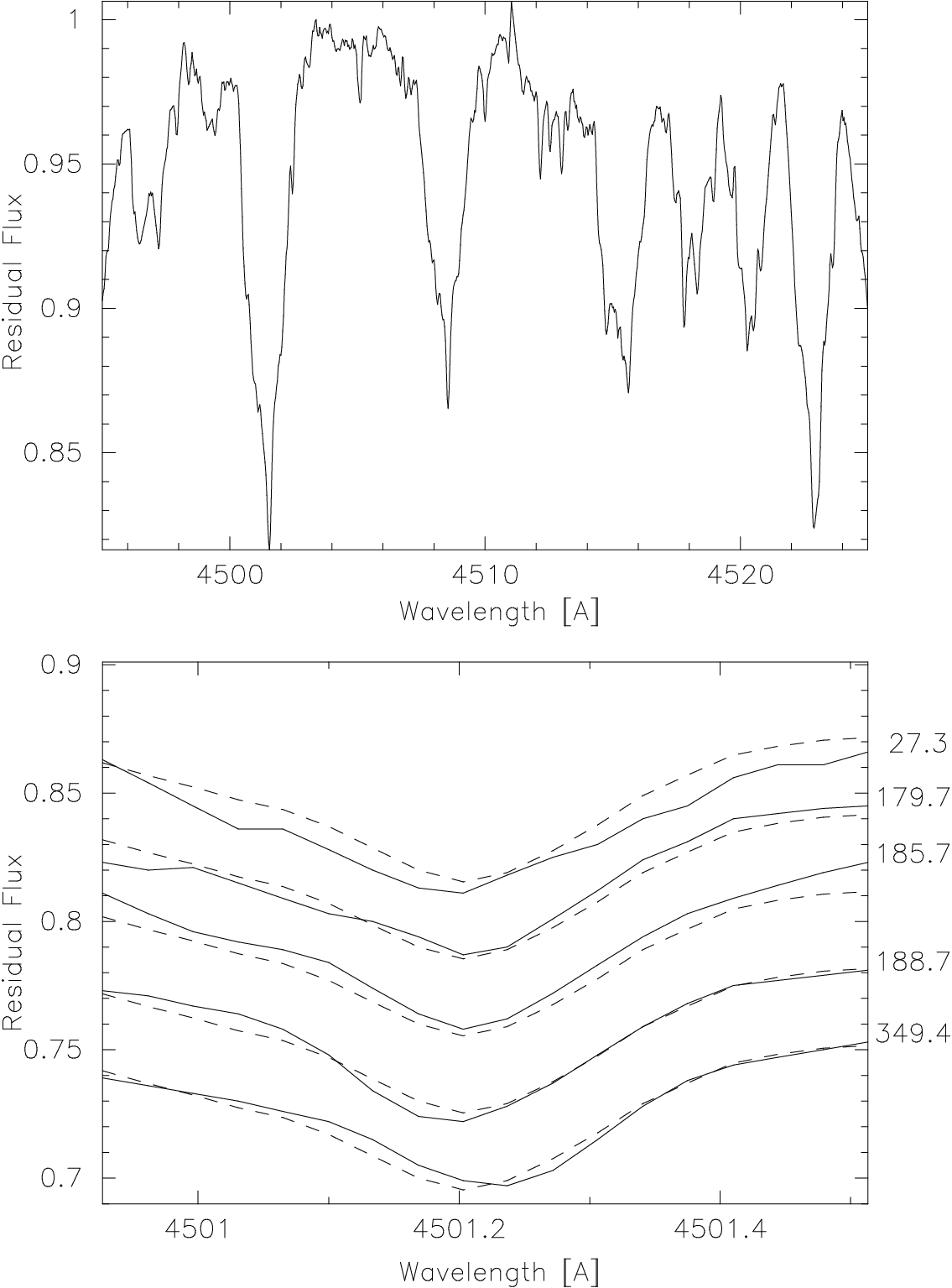}
\caption{Upper panel: a region of the composite spectrum of \object{HD\,108100} showing
both broad and narrow components. Lower panel: zoom on the narrow component showing
LPV as a function of the observation date written on the right side of the graph.
The dashed line represents the average spectrum.
}
\label{scomp}
\end{figure}

Breger et al. (\cite{bh97}) derived 2 frequencies
from a multi-site photometric campaign, confirmed by Henry \& Fekel (\cite{hf02}) 
who derived a third frequency in their Johnson data.
The spectrum is composite (Fig\,\ref{scomp}), and
we derived projected rotation velocities of 13\,\kms and 65\,\kms\
for each component,
compared to respectively 5\,\kms\ and 55\,\kms\ derived by
Henry \& Fekel (\cite{hf02}).
The Gaussian velocity fit associated with the narrow component is quite stable in our 5
spectra spread over 300\,d, varying between -6.5 and -4\,\kms, but
we have not enough data to choose between the shell model of Henry \& Fekel 
(\cite{hf02}) and the double-lined spectroscopic binary hypothesis of 
Nordstr\"om et al. (\cite{ns97}).
LPV are easily detected in the narrow component, while nothing can be said
concerning the broad one.

\subsection{\object{HD\,113867}} 

It is a prime $\gamma$\,Doradus candidate in Handler's
list (\cite{h99}), with 2 frequencies derived from the HIPPARCOS data, the first
independently confirmed by Koen (\cite{k01}).
Radial velocities are in the range [+0.5;+3.2]\,\kms.
An additional component may exist, appearing as a broad
contribution around each narrow line.
Fekel et al. (\cite{fw03}) observed such a composite spectrum, and explained it
as a shell or a double-lined spectroscopic binary.
They observed a slight change in the velocity, but within their error bars, and
rather favoured the shell hypothesis.
However, we note that compared to the preceding ones, our last spectrum,
obtained 400\,d after, is clearly the result of a Doppler shift such
as that induced by a binary motion (displacement of the whole narrow profile).
Moreover, the radial velocity measured by Fekel et al. (\cite{fw03}), 
+8.8\,\kms, is significantly out of our range.
Despite being unable to rule out the shell hypothesis, the star is certainly
a spectroscopic binary with a rather long period.
The $v\,\sin i$ value associated with the dominant component is 8.5\,\kms, 
similar to the 10\,\kms\ measured by Fekel et al. (\cite{fw03}).
LPV associated with the narrow component are represented mainly by weak relative
flux variations (also present in equivalent width variations).

\subsection {\object{HD\,211699}} 

Handler (\cite{h99}) classified this star
as a prime $\gamma$\,Doradus candidate, with 2 frequencies in the
HIPPARCOS data.
The projected rotation velocity is relatively low, 
11.6\,\kms.
LPV are very well marked, and manifest as strong
variations of the residual flux of the profiles.
Indeed, our observations represent 2 groups of data separated by
about 1 year, and the residual flux is 0.75 for the first group,
while it is around 0.55 for the second one, with a maximum variation
range of 30\,\%.
Since the FWHM of the two considered line profiles are more or less constant,
it means that the EW have strongly varied on
a 1 year time scale, increasing from 119 to 203\,m\AA\,.
The radial velocity range is [+5;+10]\,\kms.
Those associated with the group of large EW have a wider
range than the ones corresponding to low EW values.
Therefore, the status of this star is still puzzling since usually
pulsation induces moderate temperature changes (less than
100\,K) probably too small to modify significantly the level populations.
A possible explanation is stellar activity, but this latter has
been poorly reported for $\gamma$\,Doradus star candidates
(see e.g. Kaye \& Strassmeier \cite{ks98}).
Therefore this star deserves further long term observations to 
define the origin of the noted variations.

\subsection{\object{HD\,221866}} 

\begin{figure}
\centering
\includegraphics[width=8cm]{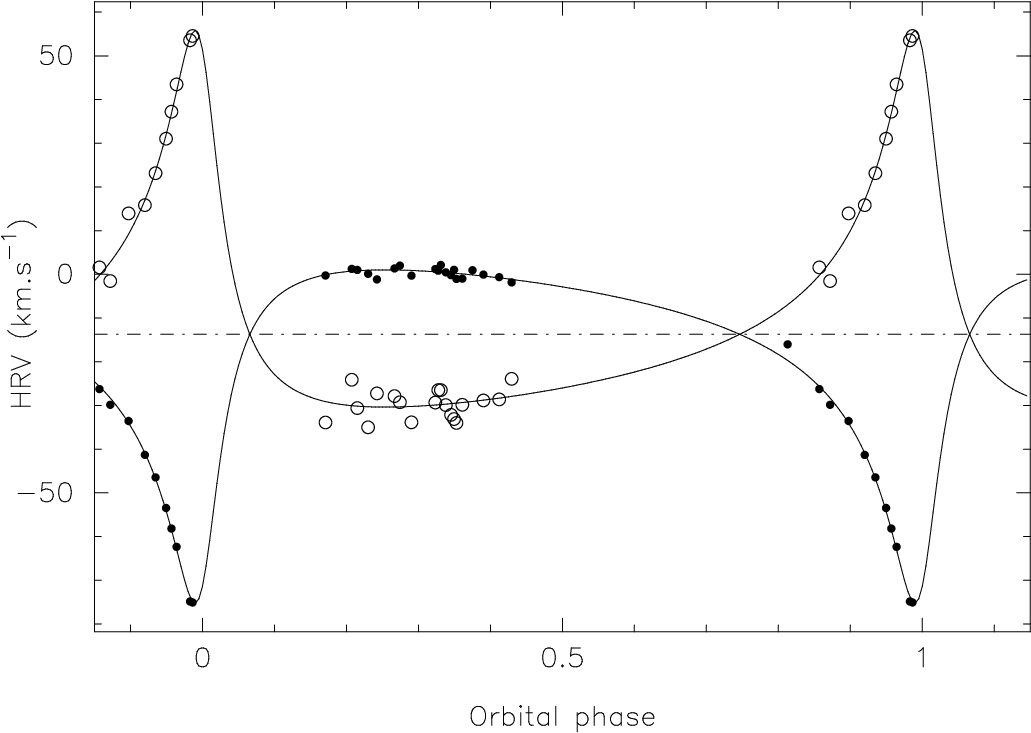}
\caption{Fit of the orbit of HD\,221866. 
Empty symbols represent the companion velocities.
The dot-dash line represents the heliocentric
velocity of the system.
}
\label{figure6}
\end{figure}

This star is cited as a prime candidate
by Handler (\cite{h99}) who derived 2 frequencies in the HIPPARCOS
data, one being common with the 3 frequencies detected by
Henry \& Fekel (\cite{hf02}) in their Johnson photometric data.
Kaye et al. (\cite{kg03}) showed that the star is a double-lined
spectroscopic binary with a primary and a secondary star of spectral
type A8m\,V and F3\,V respectively.
Our ephemeris based on our own data (Table\,\ref{table6}) confirms the results of
Kaye et al. (\cite{kg03}), and the velocity curves are
represented in Fig.\,\ref{figure6}.
\begin{table}
\caption[]{Parameters of the binary orbit for the SB2 star HD\,221866.
The final rms of the residuals is 2.28\,\kms.
}
\begin{center}
\begin{tabular}{rcl}
\hline
$P$            & = & $135.19 \pm 0.52$\,d    \\
$T_0$          & = & $2450412.41 \pm 0.29$\,d  \\
$e$            & = & $0.683 \pm 0.013$           \\
$\gamma$       & = & $-13.7 \pm 0.3$\,\kms \\
$K_1$          & = & $38.1 \pm 1.0$\,\kms \\
$\omega_1$     & = & $206.1 \pm 2.3^{\circ}$     \\
$K_2$          & = & $43.1 \pm 1.1$\,\kms \\
$\omega_2$     & = & $26.1 \pm 2.3^{\circ}$     \\
$a_1 \sin i$   & = & $(51.78 \pm 2.45)\,10^6$\,km   \\
$M_1.\sin^3 i$ & = & $1.556 \pm 0.090$\,$M_{\sun}$ \\
$a_2 \sin i$   & = & $(58.52 \pm 2.70)\,10^6$\,km   \\
$M_2.\sin^3 i$ & = & $1.376 \pm 0.080$\,$M_{\sun}$ \\
\hline
\end{tabular}
\end{center}
\label{table6}
\end{table}
The projected rotation velocities we derive are slightly lower
than the ones measured by Kaye et al. (\cite{kg03}), 
13.2 and 9.9\,\kms\ for the primary and 
the secondary, respectively.
The primary seems to present marginal LPV but the secondary
shows strong LPV, and is therefore a $\gamma$\,Doradus
star, contrary to what is suggested by Fekel et al (\cite{fw03}).
Therefore, the link between Am and $\gamma$\,Doradus stars 
suggested by Fekel et al. (\cite{fw03}) is
not yet proved.

\section{Pulsation}

Most of the stars in that sample were observed because
they present photometric variations with compatible 
$\gamma$\,Doradus frequencies.
We reanalyzed the HIPPARCOS data of our 59 stars
to derive the amplitudes associated with the main frequency.
Then, we imposed the HIPPARCOS frequency on the radial velocity
data.
The results concerning the 9 stars for which this procedure converged 
are given in Table\,\ref{table7}.
\begin{table*}
\caption[]{List of stars for which an HIPPARCOS
frequency could be deduced (3$^{\rm rd}$ column, in d$^{-1}$) with its corresponding
amplitude (4$^{\rm rd}$ column, in mmag). Then the next columns respectively
provide the corresponding radial velocity amplitude [\kms] when a sine-fit
with the $f$-frequency converges (together with the associated fraction of
the variance this frequency accounts for), and then the total velocity range 
[\kms] measured. 
Finally, last column provides the $2K/\Delta m$ amplitudes ratio 
values [\kms\,mag$^{-1}$].
}
\begin{center}
\begin{tabular}{rrlrrcc}
\hline
\multicolumn{1}{c}{HD} &
\multicolumn{1}{c}{HIP} &
\multicolumn{1}{c}{$f$} &
\multicolumn{1}{c}{a} &
\multicolumn{1}{c}{$K$} &
\multicolumn{1}{c}{Range} &
\multicolumn{1}{c}{$2K/\Delta m$}\\
\hline
\multicolumn{7}{c}{Photometric and spectroscopic fit}\\
   277 &   623      & 1.0809 & 23 & 1.13 (44\,\%) & [-9;-5]   & 49 \\
  9365 &  7280      & 1.5981 & 33 & 2.16 (86\,\%) & [-2;+4]   & 65 \\
 48271 & 32263      & 0.52436& 24 & 0.71 (10\,\%) & [-20;-15] & 30 \\
105458 & 59203      & 1.3207 & 15 & 1.44 (51\,\%) & [-14;-8]  & 96 \\
112429 & 63076      & 2.3556 & 20 & 1.84 (37\,\%) & [-15;0]   & 92 \\
113867 & 63951      & 0.93166& 18 & 0.60 (23\,\%) & [0;+3]    & 33 \\
175337 & 92837      & 1.2712 & 16 & 0.56 (70\,\%) & [-1;+1]   & 35 \\
195068 &100859      & 1.2505 & 37 & 1.69 (48\,\%) & [-29;-20] & 46 \\
211699 &110163      & 0.9328 & 42 & 1.52 (50\,\%) & [+5;+10]  & 36 \\
\hline
\end{tabular}
\end{center}
\label{table7}
\end{table*}
For completeness, Table\,\ref{table8} gives the results for the stars
observed either by HIPPARCOS or by spectroscopy.
\begin{table}
\caption[]{Same as Table\,8, but for stars for which the combined fit of
photometric and spectroscopic data is not possible.
Light amplitude [mmag] and/or radial velocity ranges [\kms] are given.
}
\begin{center}
\begin{tabular}{rrlrc}
\hline
\multicolumn{1}{c}{HD} &
\multicolumn{1}{c}{HIP} &
\multicolumn{1}{c}{$f$} &
\multicolumn{1}{c}{a} &
\multicolumn{1}{c}{Range} \\
\hline
\multicolumn{5}{c}{HIPPARCOS data fit and velocity variation} \\
  2842 &  2510      & 1.6100 & 22 & [+8;+17]  \\
  7169 &  5674      & 1.8225 & 16 & -19.4 and -9\\
 23874 & 17826      & 2.2565 & 21 & [-21;-17]\\
171244 & 90919      & 0.9964 & 14 & [-15;-13]\\
 40745 & 28434      & 1.2133 &  7 &          \\
 41448 & 28778      & 2.3815 & 18 &          \\
 65526 & 39017      & 1.5529 & 27 &          \\
 69715 & 40791      & 2.3646 & 13 &          \\
 70645 & 41488      & 1.2618 & 17 &          \\
 80731 & 46099      & 0.8964 & 29 &          \\
 86358 & 48895      & 1.2899 & 18 &          \\
 86371 & 48830      & 0.4066 & 35 &          \\
100215 & 56275      & 1.3216 & 23 &          \\
152896 & 82779      & 1.3395 & 26 &          \\
155154 & 83317      & 2.897  &  8 &          \\
160295 & 86374      & 1.3238 & 35 &          \\
167858 & 89601      & 0.76512& 41 &          \\
206043 &106897      & 2.4324 & 16 &          \\
\multicolumn{5}{c}{Only velocity variations}\\
\noalign{\smallskip}
12901  & 9807       &        &    & [+18;+22]\\
44195  & 30154      &        &    & [+9;+13]\\
49434  & 32617      &        &    & [-10;-17]\\
63436  & 38138      &        &    & [-15;-7]\\
68192  & 40462      &        &    & [+15;+19]\\
108100 & 60571      &        &    & [-6.5;+4]\\
164615 & 88272      &        &    & [-38;-33]\\
171834 & 91237      &        &    & [-29;-25]\\
171836 & 91272      &        &    &  -31.14$\pm$0.04\\
172506 & 91580      &        &    & [-40;-38]\\
174353 &            &        &    & [+9;+10]\\ 
218396 & 114189     &        &    & [-15;-8]\\
224638 & 118293     &        &    & [-4.2;-1.4]\\
224945 & 159        &        &    & [+5;+6.5]\\
\hline
\end{tabular}
\end{center}
\label{table8}
\end{table}
It appears that the velocity amplitude associated with the sine-fit is of the order of
50\,\% of the total observed range.
This, together with the fact that the fraction of the variance is sometimes
very low, can be due to several reasons. 
First, as these stars are usually multiperiodic, a single frequency
alone cannot account for the total variation.
Second, the dominant frequency is not the same in spectroscopy and in 
photometry, because spectroscopy is sensitive to higher $\ell$ degrees
than photometry.
Third, if it is the same dominant mode with the two data sets, 
the amplitude could have changed between the HIPPARCOS epoch and our spectra.
Such a change has been observed in many $\gamma$\,Doradus stars,
see e.g. Poretti et al. (\cite{pk02}). 
However, Aerts et al. (\cite{ac03}) have shown, using very
stable and homogeneous data, that such a change was not present in 
\object{HD\,48501} and only the amplitude associated with the third 
frequency was found variable in their
other $\gamma$\,Doradus star \object{HD\,12901}.
Hence, the fit of photometric and spectroscopic data cannot converge or
be unsatisfactory owing to the physical reasons described above.

Table\,\ref{table7} provides 
the $2K/\Delta m$ amplitude ratios for 9 stars.
If we exclude cases for which the fraction of the spectroscopic variance
explained by the photometric frequency is below 30\,\%, 
the amplitude ratio ranges between 35 and 96\,\kms\,mag$^{-1}$.
Actually, there are three cases: four stars (\object{HD\,277}, \object{HD\,175337},
\object{HD\,195068} and \object{HD\,211699}) have a mean value of about
40\,\kms\,mag$^{-1}$, 2 stars (\object{HD\,105458} and \object{HD\,112429})
have a mean value of 95\,\kms\,mag$^{-1}$, and one star (\object{HD\,9365})
has an intermediate value around 65\,\kms\,mag$^{-1}$. 
Of course, the HIPPARCOS photometric band is large, and
Aerts et al. (\cite{ac03}) have shown that these values were
very sensitive to the considered filter.
Nevertheless, any future theoretical description of the pulsation
in the surface layers needs to be compatible with our observational
values of these $2K/\Delta m$ amplitude ratios.

\section{Composite spectra: binarity and shell hypothesis}

Among our 59 stars, there are 27 members or potential members of
a binary system, and 21 out of these possible 27 couples are good $\gamma$\,Doradus
candidates (stars asterisked in Tab.\ref{table1} and \ref{table1bis}).
However, stars part of binary systems represent less than 50\,\% of the candidates 
of our sample.

We confirm the composite spectra noted by Fekel et al. (\cite{fw03})
and Henry \& Fekel (\cite{hf02})
for 5 stars (\object{HD\,7169}, \object{HD\,23974}, \object{HD\,108100}, 
\object{HD\,113867} and \object{HD\,160295}), and we discovered 
an additional one (\object{HD\,44333}).
All the observed photometric variations occur on a $\gamma$\,Doradus-like
timescale with an amplitude of about 20\,mmag.
\object{HD\,44333} is the only star in our sample for which no photometric
variations have been searched for so far.
Composite spectra can be interpreted as a central star in rapid
rotation surrounded by a circumstellar shell (see Fekel et al. 
\cite{fw03}).
This could be considered as an extension of the Be phenomenon towards
A-F stars.
Indeed, the known cases are very similar, because they concern
fast rotators, above 160\,\kms, 
and for the hotter A stars Balmer emission is seen (Jaschek et al. \cite{jj88}).

It was suggested that not all the lines could present narrow components, but only
the ones originating from a metastable level (Slettebak \cite{s82}).
For our 6 stars, all lines seem to be affected, therefore the shell
should reflect exactly the physical conditions of the 
embbeded stellar atmosphere.
For Be stars, the formation of an envelope around lower luminosity
stars remains a problem.
For Be stars the large rotation velocity, non radial pulsations,
activity producing large outbursts and of course radiation pressure are 
invoked, and
all these phenomena can also take place in A and F stars.
However, we note that at least two stars, \object{HD\,108100} and
\object{HD\,160295}, have
a $v\,\sin i$ of only 60\,\kms, hence much lower than the values 
derived for the stars in the
survey of Jaschek et al. (\cite{jj88}). 
Another possibility is that the shell is a remnant of the star
formation.
Indeed, one of the possibilities is that these stars are quite young,
and some of them might be related to young objects such as $\lambda$\,Bootis stars,
as supposed by Gray \& Kaye (\cite{gk99}) for \object{HD\,218396}.
But \object{HD\,218396} presents no shell features on our spectra.
In addition, the position of the confirmed $\gamma$\,Doradus stars on the
HR Diagram suggests that they can exist over a significant fraction of the
main sequence lifetime in the relevant temperature range.
The relative velocity between broad and narrow components shows that half
the stars presents an expanding shell, while the other half presents a contracting 
one, as found by Fekel et al. (\cite{fw03}).
The case of \object{HD\,160295} is puzzling, since at least 2
narrow components seem to be present. This implies, if the
shell hypothesis is valid, that 2 shells are present,
a situation encountered in e.g. RV\,Tauri stars
where it seems that multiple shell components are ejected 
through shock waves.
Finally, all but one (\object{HD\,160295}, for which we have only one 
spectrum) of our 6 stars that present composite spectra show LPV
for the narrow component (Fig.\,\ref{scomp}).
If shell variability has already been noticed (Jaschek et al. \cite{jj88}), it
concerns only its appearance/disappearance, on a timescale
of decades.
Here again, vibrations of the shell require explanation, since
pulsation modes in outer layers are mostly detected in very luminous
stars as ``strange modes''.
For these stars close to the Main Sequence, with normal luminosities,
a different mechanism has to be invoked.

Composite spectra can also be explained by the presence of
two stars of similar spectral type but with different rotation
velocities. 
The speckle technique has been able
to resolve some stars into visual binaries: \object{HD\,7169} (a close couple,
14 a.u. away; Mason et al. (\cite{mh01}), and we note that the slow
rotator component has a very stable radial velocity: s.d. 0.24~\kms), 
\object{HD\,23874} (in an eccentric 
orbit, Seymour et al. \cite{sm02}), \object{HD\,44333}
(Germain et al. \cite{gd99}, but its $\gamma$\,Dor nature is uncertain). 

Actually, for our 6 star sample, 2 are visual binaries, and 2 others 
are suspected binaries.
Our radial velocity measurements more or less confirm those
of Fekel et al. (\cite{fw03}) except for \object{HD\,113867}
where we have a shift of about 5\,\kms\ for both
components, which cannot be interpreted as due to binary motion
(but it could be an explanation if both stars orbit a third one).
The star \object{HD\,160295} could also be member of a triple system,
or the narrow line star must have a significantly cooler spectral
type to produce all the observed lines.
The relative stability of the radial velocities would imply either
a very long orbital period or systems seen almost pole-on.

Hence, both the above interpretations have problems and appear
rather as ad-hoc explanations.
In our point of view, the main problem is that only the narrow
component shows LPV.
If LPV seen in the narrow components are really related to the $\gamma$\,Doradus
frequencies, circumstellar envelope mechanisms are difficult to understand.
For this reason, the binary hypothesis, with at least one component being 
a pulsating star, seems more attractive.

\section{Conclusion}

We have presented spectroscopic observations
of 59 candidate $\gamma$\,Doradus stars detected mainly from
the HIPPARCOS space mission.
The main goal was to confirm these stars as real members of
the group through the presence of line profile variations 
typical of $g$-mode pulsations.
The $\gamma$\,Doradus stars that are confirmed by the present work,
in addition to the ``bona fide'' candidates given in Table\,\ref{table1} are 
\object{HD\,48271}, \object{HD\,70645}, \object{HD\,80731}, \object{HD\,100215}, 
\object{HD\,113867}, \object{HD\,175337}, \object{HD\,195068}.
We were unable to detect LPV in less than 40\,\% of the candidates, but most
stars being (spectroscopically) faint, the signal to noise ratios were not always
sufficient to detect very weak variations.
Moreover, for most stars we have a very limited number of spectra, so LPV
cannot be ruled out for these candidates.

In only a very few cases were we able to impose the main HIPPARCOS frequency
on the radial velocity curves deduced from the LPV.
The deduced $2K$ amplitudes are generally low (between 0.6 and 4.2\,\kms),
pointing towards a mean amplitude ratio of about 60\,\kms\,mag$^{-1}$.
The pulsation behaviour for the most interesting stars (observations are on-going)
will be described in subsequent papers.

Fekel et al. (\cite{fw03}) suggest a percentage of $\gamma$\,Doradus members of
multiple systems as high as 74\,\%. 
Our larger sample, containing however a larger proportion of stars which
are not confirmed $\gamma$\,Doradus stars, shows that this percentage seems 
to be smaller, i.e. 50\,\%.
This value is still larger than the one measured for such stars (30\,\%)
in a previous radial velocity study
(Nordstr\"om et al. \cite{ns97}).
Similar to that occuring in a number of $\delta$\,Scuti-type pulsators
(Lampens \& Boffin \cite{lb00}), we also found several $\gamma$\,Doradus
variables in binary systems with eccentric orbits.

Our sample contains 6 stars that show composite spectra. 
This behaviour can be due either to binarity or to the presence of a shell
surrounding the star.
Our data easily show that the narrow component presents LPV in 5 out
of the 6 candidates.
If a shell is really present, one has to find the mechanism that induces
LPV in this shell.
A first step would be to detect the period, if existing, of the variations of
this narrow component.

\begin{acknowledgements}
The authors thank the referee, G. Handler, who provided many useful
suggestions for improvements.
We thank the French PNPS institution for allocating a large amount of telescope time,
the necessary condition to fulfil the objectives of the programme.
AAF and JHP acknowledge DGAPA-UNAM project IN110102 for financial support.
WWW acknowledges financial support by the BM: BWK.
\end{acknowledgements}

\end{document}